%% file: manuscript.tex
\documentclass[conference]{IEEEtran} 
\IEEEoverridecommandlockouts

\usepackage[utf8]{inputenc}
\usepackage[english]{babel}

\usepackage{amsmath,amssymb,amsfonts}
\def\BibTeX{{\rm B\kern-.05em{\sc i\kern-.025em b}\kern-.08em
		T\kern-.1667em\lower.7ex\hbox{E}\kern-.125emX}}

\usepackage[final]{pdfpages}
\usepackage{multirow}
\usepackage{array}

\usepackage{graphicx} 
\graphicspath{{./figs/}}

\usepackage{color}

\usepackage[hyphens]{url}

\usepackage[acronym]{glossaries}
\loadglsentries[acronym]{acronyms}

\usepackage{enumitem}
\setlist[description]{nosep,style=multiline,leftmargin=12mm,after=\medskip}

\usepackage{amsmath}
\providecommand{\abs}[1]{\lvert#1\rvert} 
\providecommand{\norm}[1]{\left\lVert#1\right\rVert} 
\providecommand{\mean}[1]{\mathbb{E}\left\{#1\right\}} 

\newcommand\copyrighttext{%
  \textcopyright \the\year{} IEEE.
  Personal use of this material is permitted. \\
  Permission from IEEE must be obtained for all other uses, including reprinting/republishing this material for advertising or promotional purposes, collecting new collected works for resale or redistribution to servers or lists, or reuse of any copyrighted component of this work in other works.
}

\begin{document}
	
\title{On Level Crossings and Fade Durations in \\ von Mises-Fisher Scattering Channels}

\author{\IEEEauthorblockN{Kenan Turbic\IEEEauthorrefmark{1}, S\l{}awomir Sta\'nczak\IEEEauthorrefmark{1}\IEEEauthorrefmark{2}} 
	\IEEEauthorblockA{\IEEEauthorrefmark{1}
            Fraunhofer Heinrich Hertz Institute (HHI),
            Berlin, Germany \\
        }%
        \IEEEauthorblockA{\IEEEauthorrefmark{2}
            Technische Universit{\"a}t Berlin, Germany \\
            kenan.turbic@hhi.fraunhofer.de
        }%
        \thanks{This work was partially supported by the Federal Ministry for Research, Technology and Space (BMFTR, Germany) in the “Souverän. Digital. Vernetzt.” programme, joint Project 6G-RIC; project identification numbers: 16KISK020K and 16KISK030.
        It was developed within the scope of COST Action CA20120 (INTERACT).}%
}

\IEEEpubid{%
  \begin{minipage}{0.8\textwidth}\ \\[20mm] \centering
      \footnotesize \copyrighttext
  \end{minipage}
}

\maketitle

\begin{abstract}
  This paper investigates the second-order statistics of multipath fading channels with \gls{vmf} distributed scatters.
  Simple closed-form expressions for the mean Doppler shift and Doppler spread are derived as the key spectral moments that capture the impact of mobility and scattering characteristics on level crossings and fade durations.
  These expressions are then used to analyze the influence of \gls{vmf} parameters on the \gls{lcr} and \gls{afd}.
  The results show that isotropic scattering yields the highest \gls{lcr} and the lowest \gls{afd}, while fading dynamics reduce with the decreasing angular spread of scatterers.
  Moreover, mobile antenna motion parallel to the mean scattering direction results in a lower \gls{lcr} than the perpendicular motion, with the difference between the two cases increasing with the higher concentration of scatterers.
  \end{abstract}
  

\glsresetall

\IEEEpeerreviewmaketitle

\section{Introduction}
\label{sec:intro}
The angular distribution of incoming signal wavefronts is one of the key aspect in modeling wireless propagation channels.
Among various statistical models, the \gls{vmf} distribution has gained popularity due to its simplicity, flexibility, and ability to approximate complex 3D scattering environments through model mixtures \cite{Mammasis2009}.

Due to these favorable properties, the \gls{vmf} distribution has been widely adopted in the literature, e.g., \cite{Zhu2018a}-\nocite{Bian2019, Pizzo2022a}\cite{Wang2023}.
However, the complexity of the adopted channel models in these studies rendered the important channel statistics analytically intractable, limiting the analysis to numerical integration or Monte Carlo simulation approaches.
Therefore, while enabling valuable first insights, the reported studies provided limited understanding of the impact of scattering parameters, on the channel characteristics.

More recently, closed-form expressions were reported for the correlation functions \cite{Turbic2024, Zeng2024}, Doppler spectrum \cite{Turbic2025}, and \gls{lcr} and \gls{afd} \cite{Zeng2024a} in channels with \gls{vmf} scattering.
These results provided new insights into the channel characteristics associated with spatially correlated scattering.
Specifically, the simple expressions derived for a single \gls{vmf} scattering cluster in \cite{Turbic2024} and \cite{Turbic2025} reveal that the channel correlation properties and Doppler spectrum depend only on the degree of angular concentration of scatterers and on the mobile antenna motion direction relative to the mean direction of scattering, regardless of the particular azimuth and elevation angles at which the multipath cluster is observed.
On the other hand, while the \gls{lcr} and \gls{afd} are analytically treated in \cite{Zeng2024a}, the complexity of the considered mobile-to-mobile communication scenario and the convoluted form of the presented expressions obscure the explicit dependency on the scattering characteristics.

In this paper, we address this shortcoming and further investigate the level-crossing and fade duration statistics in channels with \gls{vmf} distribution of scatterers.
Based on our recently reported results for the Doppler spectrum in \gls{vmf} scattering channels \cite{Turbic2025}, we derive closed-form expressions for the spectral moments governing the second-order fading statistics, i.e., the mean Doppler shift and the Doppler spread.
The presented simple formulations explicitly capture the impact of the \gls{vmf} parameters, revealing that the temporal fading behavior is determined by the angular spread of scatterers, mobile antenna speed, and the angle between the velocity vector and the mean scattering direction.
The presented results for \gls{lcr} and \gls{afd} are practically relevant for the analysis of outage and error burst statistics, guiding the selection of transceiver parameters in practical systems \cite{Goldsmith2005book}, e.g. packet/frame length, error correction coding scheme, interleaver depth, etc.

The rest of this paper is structured as follows.
Section \ref{sec:ch_mod} presents the adopted channel model and its assumptions, Section \ref{sec:dopp_stats} derives the mean Doppler shift and the Doppler spread, while Section \ref{sec:lcr} presents expressions for the \gls{lcr} and \gls{afd}.
The obtained results are employed to analyze the impact of the \gls{vmf} distribution parameters on the \gls{lcr} and \gls{afd} in Section \ref{sec:results}, and the paper is concluded in Section \ref{sec:conclusions}.


\IEEEpubidadjcol

\section{Channel Model}
\label{sec:ch_mod}
We consider a mobile fading channel with the frequency-dependent transmission coefficient given by
\begin{align}
  H(t, f)
  &= \sum_{n=1}^{N_m} A_n e^{-j2\pi f \tau_0^n} e^{-j 2\pi f_D^n t }
  \label{eq:ch_coeff}
\end{align}
where
\begin{description}
	\item[$t$] time;
	\item[$f$] frequency;
	\item[$N_m$] number of multipath components;
	\item[$A_n$] their amplitudes;
	\item[$\tau_0^n$] initial propagation delays;
	\item[$f_D^n$] Doppler frequency shifts, i.e.,
	\begin{align}
    f_D^n
    &= \frac{1}{\lambda} \, \hat{\mathbf{k}}_n^T \mathbf{v}
    \label{eq:dopp_freq}
  \end{align}
  \item[$\lambda$] wavelength;
  \item[$\mathbf{v}$] mobile antenna velocity vector;
  \item[$\hat{\mathbf{k}}_n$] \gls{doa} unit vector, i.e.
  \begin{align}
    \hat{\mathbf{k}}_n
    = (\cos\phi_n\cos\psi_n, \sin\phi_n\cos\psi_n, \sin\psi_n)^T
    \label{eq:doa}
  \end{align}
  \item[$\phi_n$] \gls{aaoa};
  \item[$\psi_n$] \gls{eaoa};
  \item[$(\,.\,)^T$] vector transpose operation.
\end{description}
Without loss of generality, we consider a static \gls{tx} and a mobile \gls{rx}.
However, the presented model and results derived in this work also apply if their roles are reversed.

This model assumes that scattering takes place sufficiently far away from the mobile antenna, such that the planar wave propagation assumption holds and the multipath components' amplitudes, \glspl{doa} and delays are approximately constant over local areas, i.e., several tens or hundreds of wavelengths in size \cite{Molisch2011book}.
Moreover, the mobile antenna is assumed to move in a fixed direction with a constant speed.
Together, these assumptions imply that the channel is wide-sense stationary \cite{Patzold2012book}. \looseness=-1

We also assume that the number of multipath components $N_m$ is large such that, according to the central limit theorem, the channel coefficient \eqref{eq:ch_coeff} exhibits complex Gaussian distribution.
We further focus on the case when none of the components is dominant and the channel exhibits Rayleigh fading.
The local average \gls{rx} power is determined by:
\begin{align}
  \Omega
  = \sum_{n=1}^{N_m} \mean{A_{n}^2}
\end{align}
where $\mean{\,}$ denotes statistical expectation.
The slow variations in the local average power, as the mobile antenna moves across larger distances, can be represented by an independent stochastic process, i.e., large-scale fading.

The initial propagation delays in \eqref{eq:ch_coeff} represent the propagation path lengths for random mobile antenna locations at the beginning of the channel observation ($t=0$). At a fixed frequency, these differences translate to random phase shifts observed at the \gls{rx}, typically modelled by Uniformly random variables, i.e., $\varphi_0^n \sim \mathcal{U}(0, 2\pi)$.
The changes in the propagation path lengths due to mobility, under fixed \glspl{doa} and linear motion assumptions, result in time-variant phase components changing linearly in time and being perceived as apparent Doppler frequency shifts given by \eqref{eq:dopp_freq}.

The \glspl{doa} observed at a random \gls{rx} position are represented by the \gls{vmf} distribution, with the \gls{pdf} given by \cite{Mardia2000book}
\begin{align}
	p_{\phi\psi}(\phi, \psi)
  &=
  \frac{\kappa \cos\psi}{4\pi\sinh{\kappa}}
  e^{\kappa\left[ \cos\mu_\psi \cos\psi \cos(\phi - \mu_\phi) + \sin\mu_\psi \sin\psi\right]}
  , \notag \\[1mm]
  &\phantom{=}\abs{\phi} \leq \pi, \;
  \abs{\psi} \leq \pi/2
	\label{eq:vmf_pdf}
\end{align}
where
\begin{description}
	\item[$\mu_\phi$] mean \gls{aaoa};
	\item[$\mu_\psi$] mean \gls{eaoa};
	\item[$\kappa$] spread parameter.
\end{description}
The \gls{vmf} modeldescribes scattering concentrated around an arbitrary mean direction specified by the parameters $\mu_\phi$ and $\mu_\psi$, with the spread of scatterers around this direction being controlled by the parameter $\kappa$.
It includes the isotropic scattering ($\kappa = 0$) and single-point scattering ($\kappa\rightarrow\infty$) in the direction ($\mu_\phi$, $\mu_\psi$) as special cases, while accommodating arbitrary degrees of angular concentration between these two extremes. \looseness=-1

For the presented channel model, in the following section we derive the mean Doppler shift and the Doppler spread.
These two spectral moments impose the impact of the scattering distribution on the \gls{lcr} and \gls{afd}, evaluated in Section~\ref{sec:lcr}.

\section{Mean Doppler Shift and Doppler Spread}
\label{sec:dopp_stats}
The Doppler \gls{pdf} in \gls{vmf}-scattering channels was recently derived in \cite[Eq.\,17]{Turbic2025}, and can be expressed as
\begin{align}
  p_{f_D}(f)
  &=
  \frac{1}{2f_m} \frac{\kappa}{\sinh \kappa}
  e^{\kappa \, \frac{f_\mu}{f_m} \frac{f}{f_m} } \;\times
  \notag \\
  &\phantom{=}
  I_0\left( \kappa\sqrt{1 - \left(\frac{f_\mu}{f_m}\right)^2} \sqrt{1 - \left(\frac{f}{f_m}\right)^2} \right),
  \quad \abs{f}\leq f_m
  \label{eq:dopp_pdf}
\end{align}
where
\begin{description}
  \item[$f_{\mu}$] Doppler shift for the mean \gls{doa} direction, i.e.
  \begin{align}
    f_{\mu}
    = f_m (\mathbf{k}_\mu^T \mathbf{\hat{v}})
    = f_m \cos\beta
  \end{align}
  \item[$\mathbf{\hat{v}}$] motion direction unit vector, i.e., $\mathbf{\hat{v}}=\mathbf{v}/\norm{\mathbf{v}}$;
  \item[$\mathbf{k}_\mu$] mean scattering direction unit vector, i.e.
  \begin{align}
    \mathbf{k}_\mu
    = \left[ \cos\mu_\phi\cos\mu_\psi,\; \sin\mu_\phi\cos\mu_\psi,\; \sin\mu_\psi \right]^T
  \end{align}
  \item[$f_m$] maximum Doppler shift, i.e., $f_m = \norm{\mathbf{v}}/\lambda$;
  \item[$\beta$] angle between the vectors $\mathbf{k}_\mu$ and $\mathbf{\hat{v}}$;
  \item[$I_0(\,.\,)$] modified Bessel function (first kind, zeroth order).
\end{description}

By evaluating expectation with respect to this \gls{pdf}, the mean Doppler frequency is obtained as (see the Appendix)
\begin{align}
  \mu_D
  = \mean{f_D}
  &= w_\kappa f_\mu
  \label{eq:doppler_avg_vmf}
\end{align}
where
\begin{align}
  w_\kappa = \coth\kappa - \frac{1}{\kappa}
  \label{eq:w_kappa}
\end{align}
Similarly, the mean squared Doppler frequency is obtained as
\begin{align}
  \mean{f_D^2}
  &= \frac{w_\kappa}{\kappa}f_m^2 +  \left( 1 - 3\frac{w_\kappa}{\kappa} \right) f_\mu^2
  \label{eq:doppler_meansqr_vmf}
\end{align}
The Doppler spread, i.e., the standard deviation of the Doppler frequency, then follows as
\begin{align}
  \sigma_D
  &= \sqrt{\mean{f_D^2} - \mu_D^2}
  \\
  &= \sqrt{\frac{w_\kappa}{\kappa}f_m^2 +  \left( 1 - 3\frac{w_\kappa}{\kappa} - w_\kappa^2 \right) f_\mu^2}
  \label{eq:doppler_stddev_vmf}
\end{align}

For the special case of isotropic scattering ($\kappa=0$), the mean Doppler shift \eqref{eq:doppler_avg_vmf} and the Doppler spread \eqref{eq:doppler_stddev_vmf} become
\begin{align}
  \lim_{\kappa\rightarrow 0} \mu_D &= 0
  \label{eq:doppler_avg_vmf_kappa_0}
  \\
  \lim_{\kappa\rightarrow 0} \sigma_D &= \frac{1}{\sqrt{3}} \, f_m
  \label{eq:doppler_stddev_vmf_kappa_0}
\end{align}
The zero mean Doppler shift is observed due to the Doppler \gls{pdf} in this case being symmetric around zero and constant over the domain, $\abs{f_D} \leq f_m$ \cite{Turbic2025}.
On the other hand, the Doppler moments for $\kappa\rightarrow\infty$ are
\begin{align}
  \lim_{\kappa\rightarrow\infty} \mu_D &= f_\mu
  \label{eq:doppler_avg_vmf_kappa_inf}
  \\
  \lim_{\kappa\rightarrow\infty} \sigma_D &= 0
  \label{eq:doppler_stddev_vmf_kappa_inf}
\end{align}
These results are straightforwardly interpreted by considering that, in this case, scattering reduces to a single point in the direction of the mean \gls{doa}.

\begin{figure}[!t]
	\centering
	\includegraphics[width=8.6cm]{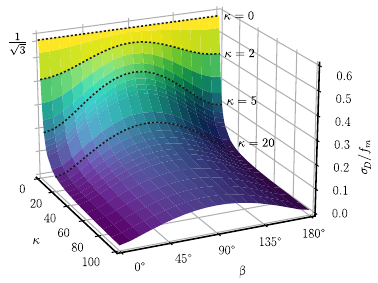}
	\caption{Doppler spread normalized by the maximum Doppler shift.}
	\label{fig:dopp_stat_stddev}
\end{figure}

The Doppler spread in \eqref{eq:doppler_stddev_vmf} is plotted in Fig. \ref{fig:dopp_stat_stddev}.
The figure shows that the Doppler spread is invariant to the motion direction for the isotropic scattering case ($\kappa=0$).
As the scattering becomes concentrated with increasing $\kappa$, the impact of the motion direction becomes apparent with the maximum Doppler spread observed for the motion perpendicular to the mean \gls{doa} direction ($\beta=90^\circ$) and the minimum one for the parallel motion ($\beta=0^\circ$, $180^\circ$).
As scattering collapses to a single point ($\kappa\rightarrow\infty$), the dependence on the motion direction is lost again as the Doppler spread asymptotically goes to zero.

Finally, we should point out that the results presented in this section apply for a single \gls{vmf}-scattering cluster.
However, they can be straightforwardly generalized to a multi-cluster scattering scenario, where the angular distribution of scatterers can be represented by a mixture of \gls{vmf} distributions \cite{Mammasis2009}.

\section{Level-Crossing Rate and Average Fade Duration} 
\label{sec:lcr}
According to \cite[Sec.~1.3]{Jakes1974ch}, the \gls{lcr} in Rayleigh fading channels can be expressed as
\begin{align}
	L_{\xi}(\rho)
	&= 2 \sqrt{\pi} \, \sigma_D \, \rho \, e^{-\rho^2}
	\label{eq:lcr_rms}
	\\
	T_{\xi}(\rho)
	&= \frac{1}{2\sqrt{\pi}} \, \frac{1}{\sigma_D} \, \frac{1}{\rho} \, \left[ e^{\rho^2}-1 \right]
	\label{eq:afd_rms}
\end{align}
where
\begin{description}
	\item[$\rho$] envelope level normalized to the \gls{rms} value, i.e., $\sqrt{\Omega}$.
\end{description}
The dependence of the \gls{lcr} on the envelope level follows from the factor $\rho \, e^{-\rho^2}$.
The maximum \gls{lcr} occurs at the level 3\,dB below the \gls{rms} value (i.e., $\rho=1/\sqrt{2}$), and is given by
\begin{align}
  L_{\xi}^{\max} = \sigma_D \sqrt{2\pi e^{-1}}
  \label{eq:lcr_rms_max}
\end{align}
The impact of the scattering characteristics follows from the Doppler spread factor, i.e., given by \eqref{eq:doppler_stddev_vmf} for \gls{vmf} scattering.
As follows from \eqref{eq:doppler_stddev_vmf} and \eqref{eq:lcr_rms}, the \gls{lcr} is directly proportional to the mobile speed (i.e., through $f_m$), while the dependence on the motion direction (i.e., through $f_{\mu}$) can be inferred from Fig. \ref{fig:dopp_stat_stddev}. \looseness=-1

For the special case of isotropic scattering ($\kappa=0$), the \gls{lcr} is obtained by replacing \eqref{eq:doppler_stddev_vmf_kappa_0} in \eqref{eq:lcr_rms}, yielding
\begin{align}
  L_{\xi}(\rho; \kappa=0)
	&= 2 f_m \sqrt{\frac{\pi}{3}} \, \rho \, e^{-\rho^2}
  \\
	T_{\xi}(\rho; \kappa=0)
	&= \frac{1}{2f_m} \, \sqrt{\frac{3}{\pi}} \, \frac{1}{\rho} \, \left[ e^{\rho^2}-1 \right]
\end{align}
Only the mobile speed is relevant in this case, while the dependance on the motion direction is lost.
This follows as the configuration of scatterers relative to the mobile antenna remains the same regardless of the direction it is moving in.
On the other hand, for the special case of single-point scattering ($\kappa=\infty$) the \gls{lcr} is zero, as follows by inserting \eqref{eq:doppler_stddev_vmf_kappa_inf} in \eqref{eq:lcr_rms}.
This result reflects the fact that no time-variant interference between multipath components occurs in this case and the \gls{rx} signal envelope remains constant.

\section{Results analysis}
\label{sec:results}
In this section, we employ the obtained analytical results to investigate the impact of the \gls{vmf} scattering parameters on the fading dynamics.
The effects of both the mean scattering direction and the degree of concentration of scatterers are analyzed. \looseness=-1

\begin{figure}[!t]
	\centering
	\includegraphics{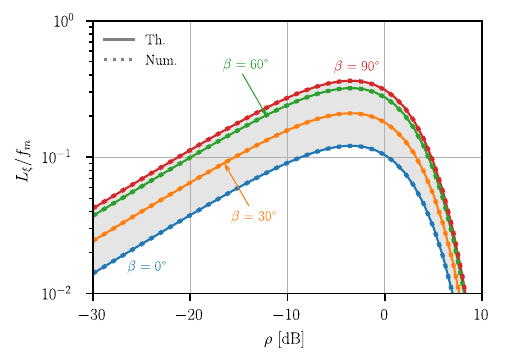}
	\caption{LCR for different motion directions relative to the mean \gls{doa} ($\kappa = 10$).}
	\label{fig:lcr_vmf_vs_beta}
\end{figure}

Fig.~\ref{fig:lcr_vmf_vs_beta} shows \gls{lcr} as a function of the (normalized) envelope level, for different relative angles between the mean \gls{doa} and the mobile antenna motion direction.
The gray-shaded area in the figure indicates the region containing the \gls{lcr} curves for all possible cases.
As one observes, motion perpendicular to the mean \gls{doa} ($\beta=90^\circ$) yields the highest \gls{lcr}, while parallel motion directly towards or away from the scattering center ($\beta=0^\circ,180^\circ$) yields the lowest rate.
We should also point out that the difference between the fading rates for these two extreme cases increases as the scattering becomes more concentrated (higher $\kappa$).

\begin{figure}[!t]
	\centering
	\includegraphics{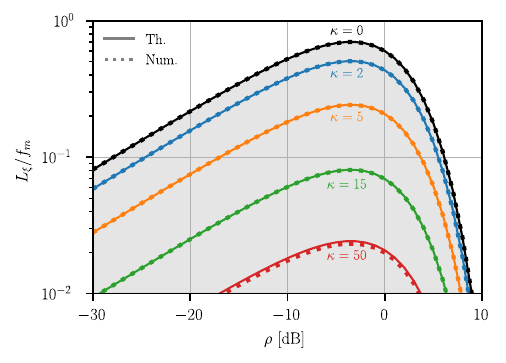}
	\caption{LCR for different levels of scattering concentration around the mean \gls{doa} ($\beta = 0^\circ$).}
	\label{fig:lcr_vmf_vs_kappa}
\end{figure}

Fig.~\ref{fig:lcr_vmf_vs_kappa} shows \gls{lcr} for different degrees of scattering concentration around the mean direction.
For the isotropic scattering case, the \gls{lcr} is the highest and independent from the motion direction.
Increasing scattering concentration is observed to result in lower crossing rates, asymptotically reducing to zero as the scatterers collapse to a single point ($\kappa\rightarrow\infty$).
This follows as the Doppler frequency shifts exhibited by different multipath components become equal and the fading becomes time-invariant, resulting in a constant signal envelope.

\begin{figure}[!t]
  \centering
  \includegraphics{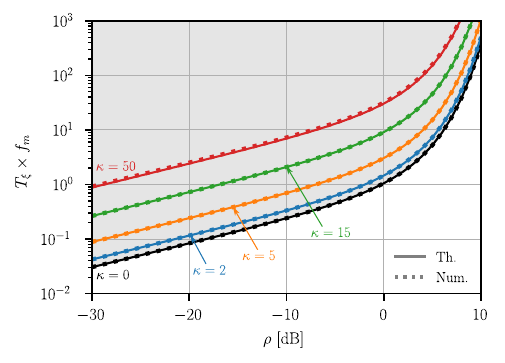}
  \caption{AFD for varying scattering concentration ($\beta = 0^\circ$).}
  \label{fig:afd_vmf_vs_kappa}
\end{figure}

For a an insight into the fade duration statistics, Fig.~\ref{fig:afd_vmf_vs_kappa} shows \gls{afd} for the same scattering parameters considered in Fig.~\ref{fig:lcr_vmf_vs_kappa}.
The isotropic scattering ($\kappa=0$) is observed to yield the shortest fade durations, with the \gls{afd} increasing with higher concentration of scattering, following the opposite trend from the \gls{lcr}.
As scattering concentrates to a single point ($\kappa\rightarrow\infty$), the \gls{afd} becomes infinite, reflecting the fact that the signal envelope for a random channel realization remains constant.

In addition to the theoretical results plotted based on the derived expressions (solid), Figs. \ref{fig:lcr_vmf_vs_beta} - \ref{fig:afd_vmf_vs_kappa} also show curves obtained by evaluating the Doppler spread via numerical integration (dotted).
The perfect agreement between the analytical and numerical results across all parameter values confirms the correctness of the derived expressions.

\section{Conclusions}
\label{sec:conclusions}
The \gls{vmf} distribution has become a popular scattering model, primarily due to its simplicity, flexibility to represent different scattering characteristics and demonstrated good fit to measurements.
However, the understanding of the impact of the \gls{vmf} scattering parameters on the channel characteristics remained largely limited until recent analytical results.

This paper presents simple closed-for expressions for the mean Doppler shift and Doppler spread in \gls{vmf} scattering channels, i.e., spectral moments that capture the effect of scattering characteristics on the second-order channel statistics.
The presented results apply to a narrowband channel transmission coefficient and to any given frequency component in the wideband channel transfer function, hence being also applicable to sub-carrier channels in \gls{ofdm} systems.

The obtained expressions are employed to analyze the impact of the mean direction and spread of scatterers on the \gls{lcr} and \gls{afd}.
The fading is observed to be slower with a lower \gls{lcr} and a higher \gls{afd} for scattering parallel to the motion direction than the perpendicular case, with the difference between the two cases being more distinct for more concentrated scattering.
The fading dynamics are the fastest for isotropic scattering, being slower for concentrated scattering and becoming zero as scattering reduces to a single point.

Finally, we should note that this work specifically focuses on simple scattering scenarios without presence of a dominant propagation path and with only \gls{tx} or \gls{rx} mobile.
The primary goal behind this choice is to better understand the fundamental impact of scattering parameters on the second-order channel statistics.
However, the extension of the presented results to channels with a dominant signal component and mobile-to-mobile scenarios is straightforward, by following the well-known procedures from the classic works in \cite{Patzold1998} and \cite{Akki1994}.

\appendix
\label{app:dopp_stats_deriv}
Here we derive the expressions for the Doppler frequency moments in \eqref{eq:doppler_avg_vmf}, \eqref{eq:doppler_meansqr_vmf} and \eqref{eq:doppler_stddev_vmf}.
To simplify the derivation, we consider the normalized Doppler frequency $\hat{f} = f/f_m$ and additionally introduce notation $\hat{f}_\mu = f_\mu/f_m$ for consistency.
By applying the simple transformation of random variables rule \cite{Papoulis2002book}, the \gls{pdf} of $\hat{f}$ follows from \eqref{eq:dopp_pdf} as
\begin{align}
  p_{\hat{f}_D}(\hat{f})
  &=
  \frac{1}{2} \frac{\kappa}{\sinh \kappa}
  e^{\kappa \, \hat{f}_\mu \hat{f} }
  I_0\left( \kappa\sqrt{1 - \hat{f}_\mu^2} \sqrt{1 - \hat{f}^2} \right),
  \quad \abs{\hat{f}}\leq 1
  \label{eq:dopp_pdf_gen}
\end{align}
We note that the relationship between the Doppler frequency moments and their normalized counterparts is given as
\begin{align}
  \mean{f_D}
  &= f_m \, \mean{\hat{f}_D}
  \label{eq:dopp_avg_vs_norm}
  \\
  \mean{f_D^2}
  &= f_m^2 \, \mean{\hat{f}_D^2}
  \label{eq:dopp_sqravg_vs_norm}
\end{align}

By applying the "Feynman technique" (derivation by parameter), the mean (normalized) Doppler shift is obtained as
\begin{align}
  \mean{\hat{f}_D}
  &= a \int_{-1}^{1} \hat{f} e^{-b \hat{f} } I_0\left( c \sqrt{1 - \hat{f}^2} \right) d\hat{f}
  \\
  &= - a \frac{\partial}{\partial b} I_{*}
  \label{eq:dopp_avg_mid}
\end{align}
where
\begin{align}
  a &= \phantom{-}\frac{1}{2} \frac{\kappa}{\sinh \kappa}
  \label{eq:dopp_avg_subs_a}
  \\
  b &= -\kappa \, \hat{f}_\mu
  \label{eq:dopp_avg_subs_b}
  \\
  c &= \phantom{-}\kappa\sqrt{1 - \hat{f}_\mu^2}
  \label{eq:dopp_avg_subs_c}
\end{align}
and
\begin{align}
  I_{*}
  &= \int_{-1}^{1} e^{-b \hat{f} } I_0\left( c \sqrt{1 - \hat{f}^2} \right) d\hat{f}
  \label{eq:I_star}
\end{align}
By employing \cite[Eqn.~6.616.5]{Gradshteyn2007book}, we obtain
\begin{align}
  I_{*}
  = \frac{2 \sinh\left( \sqrt{b^2 + c^2} \right)}{\sqrt{b^2 + c^2}}
  \label{eq:I_def}
\end{align}
By calculating the partial derivative of \eqref{eq:I_def} with respect to $b$ and replacing the result back in \eqref{eq:dopp_avg_mid}, we get
\begin{align}
  \mean{\hat{f}_D}
  &= w_\kappa \hat{f}_\mu
  \label{eq:dopp_avg_norm}
\end{align}
Replacing \eqref{eq:dopp_avg_norm} in \eqref{eq:dopp_avg_vs_norm} then yields \eqref{eq:doppler_avg_vmf}.

By using the same technique, the mean squared (normalized) Doppler frequency can be written as
\begin{align}
  \mean{\hat{f}_D^2}
  &= a \int_{-1}^{1} \hat{f}^2 e^{-b \hat{f} } I_0\left( c \sqrt{1 - \hat{f}^2} \right) d\hat{f}
  \\
  &= a \frac{\partial^2 I_{*}}{\partial b^2} 
  \label{eq:dopp_sqravg_norm_mid}
\end{align}
By evaluating the second-order derivative of \eqref{eq:I_def} in \eqref{eq:dopp_sqravg_norm_mid}, after some manipulation, we obtain:
\begin{align}
  \mean{\hat{f}_D^2}
  &= \frac{w_\kappa}{\kappa} + \left( 1 - 3 \frac{w_\kappa}{\kappa} \right) \hat{f}_\mu^2
  \label{eq:dopp_sqravg_norm}
\end{align}
Replacing \eqref{eq:dopp_sqravg_norm} in \eqref{eq:dopp_sqravg_vs_norm} then yields \eqref{eq:doppler_meansqr_vmf}.

Finally, we should point out that the $n$-th normalized Doppler frequency moment can be obtained as
\begin{align}
  \mean{\hat{f}_D^n}
  &= a \, (-1)^n \frac{\partial^n I_{*}}{\partial b^n}
  \label{eq:I_deriv_nth}
\end{align}
and the moment without normalization directly follows as
\begin{align}
  \mean{f_D^n}
  &= f_m^n \, \mean{\hat{f}_D^n}
\end{align}

\ifCLASSOPTIONcaptionsoff
  \newpage
\fi


\end{document}

%% file: acronyms.tex
\newacronym[\glslongpluralkey={Directions of Departures}]{dod}{DoD}{Direction of Departure}
\newacronym[\glslongpluralkey={Directions of Arrivals}]{doa}{DoA}{Direction of Arrival}
\newacronym[\glslongpluralkey={Angles of Departure}]{aod}{AoD}{Angle of Departure}
\newacronym[\glslongpluralkey={Angles of Arrival}]{aoa}{AoA}{Angle of Arrival}
\newacronym[\glslongpluralkey={Elevation Angles of Departures}]{eaod}{EAoD}{Elevation Angle of Departure}
\newacronym[\glslongpluralkey={Elevation Angles of Arrivals}]{eaoa}{EAoA}{Elevation Angle of Arrival}
\newacronym[\glslongpluralkey={Azimuth Angles of Departures}]{aaod}{AAoD}{Azimuth Angle of Departure}
\newacronym[\glslongpluralkey={Azimuth Angles of Arrivals}]{aaoa}{AAoA}{Azimuth Angle of Arrival}
\newacronym[first=transmitter (Tx)]{tx}{Tx}{Transmitter}
\newacronym[first=receiver (Rx)]{rx}{Rx}{Receiver}
\newacronym{vmf}{vMF}{von Mises-Fisher}
\newacronym{pdf}{PDF}{Probability Density Function}
\newacronym{cdf}{CDF}{Cumulative Distribution Function}
\newacronym{lcr}{LCR}{Level-Crossing Rate}
\newacronym{afd}{AFD}{Average Fade Duration}
\newacronym{rms}{RMS}{Root Mean Square}
\newacronym{ofdm}{OFDM}{Orthogonal Frequency Dicision Multiplex}